\documentclass[twocolumn,showpacs,preprintnumbers,amsmath,amssymb]{revtex4}
\usepackage{mathbbold}
\usepackage{graphics}
\usepackage{amsmath}
\usepackage{mathrsfs}
\usepackage{theorem}
\usepackage{overpic}
\usepackage{hyperref}
\usepackage{rotating}
\usepackage{amssymb}
\usepackage[english]{babel}
\usepackage{color,times}

\newcommand{\ket}[1]{|#1\rangle}
\newcommand{\bra}[1]{\langle #1|}

\newtheorem{theorem}{Conclusion}
\newtheorem{theorem1}{Definition}

\begin{document}
\title{Topology of Entanglement in Multipartite States with Translational Invariance}
\author{H. T. Cui}
\email{cuiht@aynu.edu.cn}
\author{J. L. Tian, C. M. Wang, and Y. C. Chen}
\affiliation{School of Physics and Electric Engineering, Anyang Normal University, Anyang 455000, China}

\begin{abstract}The topology of entanglement in multipartite states with translational invariance is discussed in this article. Two global features are foundby which one can distinguish distinct states. These are the cyclic unit and the quantised geometric phase. Furthermore the topology is indicated by the fractional spin. Finally a scheme is presented for preparation of  these types of states in spin chain systems, in which the degeneracy of the energy levels characterises the robustness of the states with translational invariance.
\end{abstract}



\maketitle

\section{Introduction}

Since the discovery of Greenberger-Horne-Zeilinger (GHZ) \cite{ghz} and W states \cite{dvc00}, it has been realised that there will exist distinct types of entanglement in multipartite states which cannot be converted into each other by stochastic local operations and classical communication (SLOCC), and thus are called SLOCC inequivalent for brevity. From the viewpoint of quantum information processing, these inequivalent entangled states show distinct abilities for the processing of quantum information. Consequently it would be interesting to find a method for the general classification of multipartite entangled states by SLOCC.

This task is  difficult due to the existence of many classifications that emphasise different features of entanglement \cite{nielson99, generalized.majorization, 4qubit, schmidt, tensor.rank, polynomial.invariants, inequality, entanglement.witness, local.unitary, topo.and.geometry, local.entanglability, lattice.gauge.approach, lorentz.invariant, majorana.representation, bastin}. This difficulty arises from the absence of Schmidt decomposition in the multipartite case. Thus information of entanglement in multipartite states cannot be obtained only by local probing of the state. However some methods have been proposed in order to achieve this goal, which focus on different aspects of multipartite entanglement. The generalisation of Schmidt decomposition to the multipartite case has been proposed \cite{schmidt}. Consequently the concept of the Schmidt tensor rank has also been introduced, for which the crucial idea is to find the minimal decomposition on a product state basis \cite{tensor.rank}. Another interesting method is that of finding  the polynomial invariant, which is composed from the superposition coefficients of a state on a preferred product state basis, and thus  is invariant under SLOCC. The distinct patterns of polynomial invariants can be used to classify the multipartite entanglement \cite{polynomial.invariants}. In addition there are extensive efforts in other areas including generalised majorisation \cite{generalized.majorization}, criteria based on inequalities \cite{inequality}, entanglement witnesses \cite{entanglement.witness}, and local unitary equivalence \cite{local.unitary}.

A common feature of previous studies is that  local measurements  are applied to obtain information on entanglement in multipartite states. However this method becomes tedious for an increasing number of particles. Thus it is interesting to characterise the  entanglement of states from a global viewpoint which is invariant under SLOCC and independent of local features in the states. Geometric or topological descriptions of entanglement are therefor proposed  \cite{topo.and.geometry}. In this way, distinct types of  entanglement can be indicated directly by different geometric or topological features. Unfortunately this method also becomes inefficient with increases in the number of particles. Recently a method for the classification of entanglement in states using global symmetry has been proposed \cite{lattice.gauge.approach, lorentz.invariant, majorana.representation, bastin}. In this way, some novel structures in multipartite states are revealed, which can be used to identify distinct types of entanglement. For example, a complete classification of multipartite entangled states with permutational invariance has been performed  by determination of the distributions of the roots of the Majorana polynomial \cite{majorana.representation} or the permutational unit in the states \cite{bastin}. Furthermore experimental schemes for differentiating  distinct types of entanglement are also proposed \cite{experimental.scheme}. Although this discussion is abstract, it has physical interest since  states with this symmetry can be prepared readily in realistic physical systems \cite{pqsv11}.

However, from the viewpoint of preparation of entanglement in concrete systems, especially in condensed matter systems, translational symmetry would be more applicable. Thus in this article we  focus on differentiating multipartite entangled states with translational invariance (TI). This idea was first put forward in our previous work \cite{cui10}, and we would like to present a systemic discussion in this article. Our study shows that two global features exist for states with TI by which one can distinguish distinct states of TI. Furthermore we show by a spin chain model that these states correspond to its energy levels. Additionally distinct states of TI will have different energies, by which one can distinguish them. More importantly the topology of states of TI can also be demonstrated by this picture.  Thus the classification of states of TI is meaningful from an experimental viewpoint. It should be noted that this idea is compatible with  recent findings \cite{quantum.phase} showing that different quantum phases in condensed matter systems will display differing computational power. Conversely this result implies that different quantum phases would correspond to distinct types of quantum entanglement \cite{wen02}.

\section{Translational invariance of states and the topological features}
The \emph{symmetry of a state} is defined in the following way:
\begin{theorem1}
For a symmetry operator $\hat{S}$, if
\begin{equation}\label{ss}
\hat{S}\ket{\psi}=c\ket{\psi},
\end{equation}
then $\ket{\psi}$ is said to have symmetry $S$ or be invariant under $\hat{S}$.
\end{theorem1}
Within this definition $c$ is complex with $|c|=1$, and $\hat{S}$ is defined in a finite dimensional Hilbert space for mathematical rigority. Because $\hat{S}$ is a normal operator, $\ket{\psi}$s with different $c$ are orthogonal to each other, and a basis of Hilbert space with symmetry $S$ can generally be constructed. 

\subsection{Translational invariance of the state and cyclic unit}
\begin{figure}[t]
\center
\includegraphics[bbllx=31, bblly=32, bburx=584, bbury=491, width=5cm]{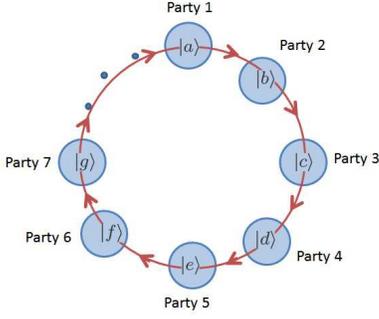}
\caption{Diagram of the translation operation $\hat{T}$ of $N$-qubit states with periodic boundary conditions. The ket vectors denote distinguishable single party states.  Then $\hat{T}$ corresponds to a clock wise rotation of all single party states in order. }
\label{ti}
\end{figure}

Focusing only on the one-dimensional case for simplicity, the translation operator $\hat{T}$ of a state corresponds to a global moving of all single party states in order. Single party state refers to an individual particle or party, and these are distinguished by arabic numbers. With periodic boundary condition (PBC) of the single party states $\ket{\phi}_{N+i}=\ket{\phi}_i (i=1,2,\cdots, N)$, where $N$ is the number of parties,  $\hat{T}$ corresponds to the clockwise cyclic permutation of all single party states as displayed in Fig.\ref{ti}. The crucial feature  is that the order of single party states is kept unchanged. Consequently, the \emph{cyclic unit} can be defined as the sequence of single party states $\left\{\ket{a}, \ket{b}, \cdots\right\}$, which uniquely determines the state. In the appendix, several examples of states of TI are presented. For example, the cyclic unit of $\ket{W_1}_3=\tfrac{1}{\sqrt{3}}\left(\ket{100}+\ket{010}+\ket{001}\right)$, known as the standard $W$ state,  is $\left\{\ket{1}, \ket{0}, \ket{0}\right\}$, and obviously  $\ket{W_1}_3$ is just the superposition of all possible cyclic permutations of single party states in the cyclic unit. By this characterisation $\ket{W_1}_3$ is then TI. More examples can be found in the appendix. For instance, the cyclic unit of $\ket{W_1}_4=\tfrac{1}{2}\left(\ket{1000}+\ket{0100}+\ket{0010}+\ket{0001}\right)$ is $\left\{\ket{1}, \ket{0}, \ket{0},\ket{0}\right\}$, while it is $\left\{\ket{1}, \ket{1}, \ket{0},\ket{0}\right\}$ for $\ket{W_3}_4=\tfrac{1}{2}\left(\ket{1100}+\ket{0110}+\ket{0011}+\ket{1001}\right)$. The two states have different cyclic units, and  are therefore  SLOCC inequivalent  because of translational symmetry of state. (This will not apply when a state of TI is composed from two or more cyclic units. Hence the following proof is restricted to the case that there is one and only one cyclic unit in the state of TI).

This point can be proved formally as follows. $\ket{\psi_1}$ and $\ket{\psi_2}$ are two arbitrary states with different cyclic units, denoted as $S_1$ and $S_2$.  In general $S_2=\otimes_{i=1}^N M_i S_1$, where $M_i$ denotes a local operation imposed on the $i$-th party. The meaning of "different" is  that one cyclic unit cannot be converted into another by the isotropic transformation $M^{\otimes N}$, i.e., $S_2\neq M^{\otimes N} S_1$. By the definition of the cyclic unit, one has (without normalisation)
\begin{eqnarray}
\ket{\psi_1}&=&\sum_{n=0}^{N-1} \hat{T}^n \ket{S_1}, \nonumber\\
\ket{\psi_2}&=&\sum_{n=0}^{N-1} \hat{T}^n\ket{S_2}=\sum_{n=0}^{N-1} \hat{T}^n\left(\otimes_{i=1}^NM_i \ket{S_1}\right),
\end{eqnarray}
where we have supposed that $\ket{\psi_{1(2)}}$ are the eigenstates of $\hat{T}$ with $c=1$, i.e., $\hat{T}\ket{\psi_{1(2)}}=\ket{\psi_{1(2)}}$. As for other values of $c$, one always finds a local unitary operation to transform the state into the form with $c=1$, and meanwhile keeps the cyclic unit unchanged as indicated  in the  appendix. Ultimately it corresponds to justify the equivalence of $\ket{\psi_1}$ and $\ket{\psi_2}$ since the local unitary operation cannot change the SLOCC equivalence of the states.

Suppose $\ket{\psi_2}$  SLOCC equivalent to $\ket{\psi_1}$, then $\ket{\psi_2}=\otimes_{i=1}^N O_i \ket{\psi_1}$, in which $O_i$ denotes the local operation imposed on the $i$-th party. Then
\begin{eqnarray}
&&\hat{T}\ket{\psi_2}=\ket{\psi_2}\nonumber\\
&\Rightarrow& \hat{T} \otimes_{i=1}^N O_i \ket{\psi_1}\nonumber\\
&=& \hat{T} \otimes_{i=1}^N O_i \hat{T}^{-1} \hat{T} \ket{\psi_1}\nonumber\\
&=& \hat{T} \otimes_{i=1}^N O_i \hat{T}^{-1} \ket{\psi_1}\nonumber\\
&=& \otimes_{i=1}^N O_i \ket{\psi_1}\nonumber\\
&\therefore& \hat{T} \otimes_{i=1}^N O_i \hat{T}^{-1}=\otimes_{i=1}^N O_i
\end{eqnarray}
The result above implies that $\otimes_{i=1}^N O_i$ is also translationally invariant, i.e., $\otimes_{i=1}^N O_i=O^{\otimes N}$. Consequently
\begin{eqnarray}
O^{\otimes N}\ket{\psi_1}&=&\sum_{n=0}^{N-1} O^{\otimes N} \hat{T}^n \ket{S_1}\nonumber \\
&=&\sum_{n=0}^{N-1} \hat{T}^n O^{\otimes N} \ket{S_1}\nonumber\\
&=&\ket{\psi_2}
\end{eqnarray}
Then $S_2=O^{\otimes N}S_1$, which is obviously contrary to the original presumption.




Conversely if $\ket{\psi_1}$ and $\ket{\psi_2}$ are SLOCC inequivalent, they must have different cyclic units. The proof of this is as follows. Assuming that $\ket{\psi_1}$ and $\ket{\psi_2}$ have the "same" cyclic units, i.e. $S_2=M^{\otimes N}S_1$, and so $\ket{\psi_2}=\sum_{n=1}^N \hat{T}^n\ket{S_2}=\sum_{n=1}^N \hat{T}^nM^{\otimes N}\ket{S_1}$. it can be noted that $\hat{T}^nM^{\otimes N}=M^{\otimes N}\hat{T}^n$, and $\ket{\psi_2}= M^{\otimes N}\sum_{n=1}^N\hat{T}^n\ket{S_1}=M^{\otimes N} \ket{\psi_1}$. This result is obviously contrary to the prerequisite that the two states are SLOCC inequivalent. Thus the proof is completed.

It should be pointed out that the discussion above can be generalised to any arbitrary number of parties. For simplicity and clarity, the following discussion is restricted to the qubit case.

\subsection{Quantized Geometric Phase}

An interesting feature is that a phase difference  will exist between two adjacent permutations of a cyclic unit, as if the system was penetrated by a flux. Consequently a global phase factor can be identified by $\hat{T}$. Noting that $\hat{T}$ with PBCs actually corresponds to a planar rotation group $C_N$, we then obtain the first conclusion:
\begin{theorem}\label{c1}
   In an N-party Hilbert space, for the translational operation $\hat{T}$, $c$ can take the  values $e^{i\frac{2n\pi}{N}}$ where $n=0,1,2,\cdots, N-1$.
\end{theorem}
Actually the values of $c$ constitute the character table of the symmetry group $C_N$ \cite{cornwell}. In addition,  since $\hat{T}$ corresponds to an adiabatic transformation  in state space, $c$ has geometric meaning as it is a geometric phase factor \cite{berry84}. Conclusion \ref{c1} implies  that the value of the geometric phase is quantised, and thus there exists a nontrivial topological structure in state space. This point can be demonstrated by the observation that the $N$ eigenstates with different $c$ actually characterise the statistics  of a quasi-particle with fractional spin $1/N$; shen one quasi-particle transforms circularly, it will acquire a phase $2\pi/N$. 


Furthermore it is known from group theory \cite{cornwell} that there exists $n$-th and $m$-th rotational axes for the planar rotation group $C_N$ if $N=n\times m (n,m\neq 1)$. As for states of TI, this means that the cyclic unit itself will have underlying periodic structure. Then the value of $c$ is determined by the character of $C_n$ and $C_m$ together. Several examples are presented in the appendix. For the 4 qubit case, there are states $\ket{\text{GHZ}'_{1(2)}}_4$ with the cyclic unit $\left\{\ket{1}, \ket{0}, \ket{1}, \ket{0}\right\}$,  composed of two length-two character stings "10". Thus we call these states $2$-periodic. Similarly for the 6 qubit case there are states $\ket{W^{(\prime)}_{0}}_6, \ket{T^{(\prime)}_{0}}_6, \ket{T^{(\prime)*}_{0}}_6$  with the cyclic unit $\left\{\ket{1}, \ket{0}, \ket{0}, \ket{1}, \ket{0},\ket{0}\right\}$, composed of two length-three character stings "100". These states are then called 3-periodic. The \emph{$n$-periodic} state of TI is defined as that for which the corresponding cyclic unit is composed from the sting with minimal length $n$. The 1-periodic states can also be defined in this manner as $\ket{11\cdots 1}$ and $\ket{00\cdots 0}$, which are obviously SLOCC equivalent to fully separable states. Actually the periodicity of the states determines the minimal  permutations needed for the cyclic unit to return back to its original form.

We then have the second conclusion:
\begin{theorem}\label{c2}
For a $N=n\times m$ qubits system, there exists $n$-periodic or $m$-periodic states of TI, for the periodicity is determined completely by the minimal length of  substructure in the cyclic unit.
\end{theorem}
From the viewpoint of topology, this conclusion implies that there could exist distinct quasi-particles with different fractional spins in the same state space. Consequently the periodicity of a state is also a demonstration of a topological feature, in state of TI, and this can also be indicated  by the quantised geometric phase.

So far we have defined two global features of states of TI, the cyclic unit and quantized geometric phase factor. Generally for a specific cyclic unit there exists many states with different geometric phase factors, and  a specific geometric phase factor corresponds to many states with different cyclic units. So one has to combine the two features together in order to differentiate individual states.

\section{Complete Hilbert space basis of states of TI and the APPLICATION TO CONCRETE SYSTEMS}

For an $N$-qubit system, one can find a complete orthogonal Hilbert space basis of states of  TI, as shown in the appendix for up to 6 qubits. This point can be demonstrated as follows. The states $\ket{00\cdots 00}$ and $\ket{11\cdots 11}$ are trivially of TI in themselves. For any other computational basis state (a product state consisting of 0s and 1s) with period $k$, one can always find its $k-1$ other cyclic states by applying  $\hat{T}$. Then by introducing a phase difference $2m\pi/k(m=0, 1,2, \cdots k-1)$  between these cyclic states, one can obtain $k$ states of TI. Conclusively $k$ computational states, related by $\hat{T}$, can construct $k$ states of TI. Then $2^N$ basis states of TI can be found. With this algorithm, the cyclic unit corresponds to the computational basis state, and can be found  readily. In the appendix the complete basis of states of TI up to six qubits have been presented, where the basis states are classified into several sets by the cyclic unit and quantised geometric phase.

Interestingly the basis states of TI corresponds to the energy levels of a spin-half chain with ferromagnetic interaction, as shown in Table \ref{t1}, which means that the basis states of TI could be realised in a concrete system. 
Moreover the topology of states of TI can also be demonstrated by the robust degeneracy of energy levels, since the breaking of degeneracy needs a the long-range interaction, i.e. $H_1$ or $H_2$ in Table \ref{t1}. Now the underlying physics of the cases  in this table are clarified in order.


\emph{3-qubit case}. There are two degenerate ground states, $\ket{1}_3$ and $\ket{0}_3$, which are fully separable. When the state of one or two qubits is flipped, one obtains the first excited states $\ket{W_{1(2)}}_3, \ket{T_{1(2)}}_3$ and $\ket{T^*_{1(2)}}_3$, which are 3-periodic and are equivalent to each other as shown in the appendix. Physically the six degenerate states characterise a fractal spin $1/3$ Moreover, since there is not any symmetry breaking, the degeneracy is  protected by translational symmetry \cite{symmetry.protected.phase}, and thus is stable against local perturbations without destroying TI. From this viewpoint the excitation is topologically nontrivial.

In fact these excited states are the so called chiral spin states, first discussed in reference \cite{wen89}. The chirality can be displayed by the polarisation of the states;  for example, $\hat{T}\ket{T_1}_3=e^{- i 2\pi/3}\ket{T_1}_3$, while $\hat{T}\ket{T_2}_3=e^{i 2\pi/3 }\ket{T_2}_3$. Whereas upon reversing the orientation of the cyclic permutation, then $\hat{T}^{-1}\ket{T_1}_3=e^{i 2\pi/3 }\ket{T_1}_3$ and $\hat{T}^{-1}\ket{T_2}_3=e^{-i 2\pi/3 }\ket{T_2}_3$.  This point implies that $\ket{T_1}_3$ and $\ket{T_2}_3$ have opposite orientations, related to the cyclic permutation, and thus are polarised in themselves.  A similar picture is found for $\ket{T^*_1}_3$ and $\ket{T^*_2}_3$. Consequently $\ket{W_1}_3$ and $\ket{W_2}_3$ are non-polarised by this reasoning. The two distinct types of states can be distinguished by introducing the phase-detuning interaction
\begin{equation}\label{hp}
H'=-\tfrac{1}{2}\sum_{n=1}^3\left( e^{i\phi}\sigma^+_n\sigma^-_{n+1}+e^{-i\phi}\sigma^-_n\sigma^+_{n+1}\right),
\end{equation}
for which the chiral spin states have different energies by the choice of the value of $\phi$ . For instance $\ket{T_1}_3$ has the minimal energy when $\phi=-2\pi/3$. Whereas when $\phi=2\pi/3$, the energy of  $\ket{T^*_1}_3$  is the minimal, and when $\phi=0$, $\ket{W_1}$ has the minimal energy. Accordingly one can distinguish these polarised states in experiments by choosing $\phi$ properly, which can be realised by adding a proper flux.

\emph{4-qubit case}. There are two degenerate ground states $\ket{1}_4$ and $\ket{0}_4$. Two different cases can be found in the first excited level; the first case occurs when the state is flipped only for a single qubit: $\ket{W_{1(2)}}_4, \ket{T_{1(2)}}_4, \ket{T^*_{1(2)}}_4$ and $\ket{T'_{1(2)}}_4$. The second is that two consecutive qubits are flipped simultaneously: $\ket{W_3}_4, \ket{T_3}_4, \ket{T^*_3}_4$ and $\ket{T'_3}_4$. The two situations can be distinguished by introducing the next nearest neighbor perturbation $H_1$.  The two highest energy states are $\ket{\text{GHZ}'_{1(2)}}_4$ , which are  equivalent to $\ket{\text{GHZ}_{1(2)}}_4=\frac{1}{\sqrt{2}}\left(\ket{1111}\pm\ket{0000}\right)$.

The first excited states consist of two different types of cyclic units.  Although the two types of excited states are inequivalent, they both characterise the same topological state with fractional spin $1/4$. This point implies that this situation will be more stable that that of the 3-qubit case since it has a higher level of degeneracy. There exists also 2-periodic states of TI: $\ket{\text{GHZ}'_{1(2)}}_4$. Thus two different topologies can appear in 4-qubit system.

Similarly to the case of 3 qubits, the differently polarised states in the first excited level can be distinguished by a generalized $H'$. However in this case, our calculations show that $H'$ would mix $\ket{\text{GHZ}'_{1(2)}}_4$ and the first excited level. Hence $H'$ cannot be used to distinguish the higher-level excited states.

\emph{5-qubit case.} Since 5 is a prime number, this case displays no special topological features.

\emph{6-qubit case.} The first excited states can be divided into three classes C4, C5 and C7 (see the appendix for their meanings), which correspond to the situations that the states are flipped  for one, two and three consecutive qubits respectively as shown in the appendix. By introducing the nearest neighbour  interaction $H_1$, one can distinguish C4 from the other two cases. Then an additional nearest neighbour interaction $H_2$ is necessary to distinguish the C5 and C7 classes. A similar situation can be found for the second excited states, which can be divided into three distinct classes, C3, C6 and C8. The highest energy levels are $\ket{\text{GHZ}'_{1(2)}}_6$, which are equivalent to $\ket{\text{GHZ}_{1(2)}}_6=\frac{1}{\sqrt{2}}\left(\ket{111111}\pm\ket{000000}\right)$.

The second excited level includes two topological states with fractional spins $\frac{1}{6}$ and $\frac{1}{3}$ respectively. In addition the highest energy level belongs to the topological state with half spin. Thus there are three types of topologically distinct excitations in the 6-qubit system.


We have demonstrated that the basis states of TI can be the energy levels in spin chain systems. Seemingly this point means that from the viewpoint of quantum entanglement, that $H_i(i=0,1,2)$ could used as an entanglement indicator. However the situation is not so, as shown below.

The definition of an entanglement indicator is $W_{ent}=\mathrm{tr} \left[- \rho_{ent} H_i\right]- E_{sep}$, in which $E_{sep}=\min_{\rho_{sep}\in \mathcal{S}}\mathrm{tr} \left[- \rho_{sep} H_i\right]$ and $\mathcal{S}$ is the set of fully separable states \cite{eneryentangl}. If $\rho_{ent}$ is entangled, $W_{ent}\leq 0$; while if it is fully separable, $W_{ent}\geq 0$. Consider the fully separable state of TI $\ket{\psi_{sep}}_N=\left(\sqrt{z_1}\ket{1}+\sqrt{z_0}\ket{0}\right)^{\otimes N}$, in which $z_0$ and $z_1$ are complex numbers with the constraint $\left|z_1\right|+ \left|z_0\right|=1$. Then one can find that $_N\bra{\psi_{sep}} (-H_i)\ket{\psi_{sep}}_N = N \left(\left|z_1\right|- \left|z_0\right|\right)^2$. With respect to the subject of this article, we limit $\rho_{sep}$
to be of TI. Consequently the calculation shows that $E_{sep}=0$. As shown in Table \ref{t1}, there are some entangled states that do not satisfying this criterion.

As for mixed states, the result is similar. For example, consider the Werner-like state $\rho_W=\tfrac{1-p}{2^N}\mathbbold{1}_{2^N}+p\ket{\psi}\bra{\psi}$, in which $\ket{\psi}$ denotes the basis state of TI. It is easy to find $\mathrm{tr} \left[- \rho_W H_i\right]=p\bra{\psi} (-H_i)\ket{\psi}\geq 0$ . However, it is known that the Werner state of two qubits is entangled when $p>1/3$ \cite{werner}.

The counter-examples show that $H_i$ cannot be used as an entanglement indicator. Actually we use $H_i$ just to demonstrate the topology of the basis states of TI and their realisation in concrete systems. It is still difficult to find a general entanglement indicator for these states, although they share some common features.

\section{Further Discussion and Conclusions}

Since any $N$-qubit state can be written as a superposition of  basis states of TI, two situations can occur; if the state is a superposition of   two or more basis states of TI with the same geometric phase factor $c$, the state is also of TI, and if not, it must be not of TI. However, in both cases, one cannot determine the topology of the state only by the superposition form, since the superposition of the  basis states of TI could also be fully separable, i.e., topologically trivial, for example $\ket{\psi_{sep}}_N$ and $\ket{100}=\tfrac{1}{\sqrt{3}}\left(\ket{W_1}_3+\ket{T_1}_3+\ket{T^*_1}_3\right)$.

A special case are the $N$-qubit GHZ states, which are obviously permutationally invariant and thus 1-periodic by our definition. Although they are also the ground states of $H_0$, one can introduce the non-local interaction to break the degeneracy, as shown in Table \ref{t2} \cite{ghz}. The unique preparation of a GHZ-like state in a concrete system requires $m$-body coupling with $m\geq(N+1)/2$ \cite{ghz}. This nonlocal interaction is a great challenge for current experiments. Extensive studies can also be found in references \cite{majorana.representation, bastin, ghzsym}.

In conclusion, two topological features of states of TI, the cyclic unit and quantized geometric phase, has been identified in this article. Consequently the underlying topology of the states can be disclosed by the two features. Moreover the topology could be manifested physically by a fractional spin. In addition the state degeneracy for a specific cyclic unit reflects the robustness of the topology with respect to local perturbation without symmetry breaking. By a spin chain model, we show that the topological features can emerge from  excitation above fully separable ground states. Thus it is possible to identify the topological features in experiments.

Although our discussion is limited to states of TI, it has extensive interest. Especially it is of great interest  to discuss the quantum statistics of the fractional spin and its possible connection to the fractional excitations in quantum Hall systems \cite{hall}, in which the symmetry of translation also plays an important role for the construction of Landau levels \cite{ti}. From another viewpoint, the physical states are only a tiny set in the whole Hilbert state space \cite{pqsv11}, and have to display some symmetry in order that they can be realiszed in concrete systems. Thus we hope that this work will provide a new understanding of quantum entanglement.

\begin{acknowledgements}
H.T.C. acknowledges fruitful discussion with Dr. Chang-Shui Yu. This work is supported by NSF of China, Grant No. 11005002 (Cui) and 11005003 (Tian), New Century Excellent Talent of M.O.E (NCET-11-0937), and Sponsoring Program of Excellent Younger Teachers in universities in Henan Province of China (2010GGJS-181).
\end{acknowledgements}

\newpage

{\color[rgb]{0.99, 0, 0}

\begin{widetext}
\begin{table}
\center
\begin{tabular}{c||p{4.5cm}|c|c|c}
\hline\text{No. of qubit} &  \text{state}  &  $H_0=-\sum_{n=1}^{N}\sigma^z_n\sigma^z_{n+1}$  &  $H_1=-\sum_{n=1}^{N}\sigma^z_n\sigma^z_{n+2}$  &  $H_2=-\sum_{n=1}^{N}\sigma^z_n\sigma^z_{n+3}$ \\
\cline{1-5}
 &$\ket{1}_3, \ket{0}_3$ & -3 & - & - \\ \cline{2-5}
 \raisebox{1ex}[0pt]{3-qubit}&$\ket{W_{1}}_3, \ket{T_{1}}_3, \ket{T^*_{1}}_3$ \linebreak $\ket{W_{2}}_3, \ket{T_{2}}_3, \ket{T^*_{2}}_3$ & 1 & - & -\\ \cline{1-5}

&$\ket{1}_4, \ket{0}_4$ & -4 & - & - \\ \cline{2-5}
 &$\ket{W_{1}}_4, \ket{T_{1}}_4, \ket{T^*_{1}}_4,\ket{T'_{1}}_4$ \newline $\ket{W_{2}}_4, \ket{T_{2}}_4, \ket{T^*_{2}}_4,\ket{T'_{2}}_4$ & 0 & 0 & -\\ \cline{2-5}
& $\ket{W_{3}}_4, \ket{T_{3}}_4, \ket{T^*_{3}}_4, \ket{T'_{3}}_4$&0 & 4 & -\\ \cline{2-5}
\raisebox{5.5ex}[0pt]{4-qubit}&$\ket{\text{GHZ}'_1}_4, \ket{\text{GHZ}'_2}_4$&4&-&- \\ \cline{1-5}

&$\ket{1}_5, \ket{0}_5$ & -5 & - &- \\ \cline{2-5}
&$\ket{W_{1}}_5, \ket{T_{1}}_5, \ket{T^*_{1}}_5,\ket{T'_{1}}_5,\ket{T^{'*}_{1}}_5$ \newline  $\ket{W_{2}}_5, \ket{T_{2}}_5, \ket{T^*_{2}}_5,\ket{T'_{2}}_5,\ket{T^{'*}_{2}}_5$ & -1  &  -1 & -\\ \cline{2-5}
&$\ket{W_{3}}_5, \ket{T_{3}}_5, \ket{T^*_{3}}_5,\ket{T'_{3}}_5,\ket{T^{'*}_{3}}_5$ \newline  $\ket{W_{4}}_5, \ket{T_{4}}_5, \ket{T^*_{4}}_5,\ket{T'_{4}}_5,\ket{T^{'*}_{4}}_5$ & -1  &  3 & -\\ \cline{2-5}
\raisebox{6.5ex}[0pt]{5-qubit}&$\ket{W_{5}}_5, \ket{T_{5}}_5, \ket{T^*_{5}}_5,\ket{T'_{5}}_5,\ket{T^{'*}_{5}}_5$ \newline  $\ket{W_{6}}_5, \ket{T_{6}}_5, \ket{T^*_{6}}_5,\ket{T'_{6}}_5,\ket{T^{'*}_{6}}_5$ & 3  &  -1 & -\\ \cline{1-5}

&$\ket{1}_6, \ket{0}_6 $& -6 & - &- \\ \cline{2-5}
&$\ket{W_{0}}_6, \ket{T_{0}}_6,\ket{T_{0}}_6$ \newline
$\ket{W^{\prime}_{0}}_6, \ket{T^{\prime}_{0}}_6,\ket{T^{\prime*}_{0}}_6$ & 2 & 2 & -6 \\ \cline{2-5}
&$\ket{W_{1(2)}}_6, \ket{T_{1(2)}}_6,\ket{T^*_{1(2)}}_6, $\newline$ \ket{T'_{1(2)}}_6, \ket{T'^*_{1(2)}}_6, \ket{T''_{1(2)}}_6 $& -2 & -2 & -2 \\ \cline{2-5}
&$\ket{W_{3(4)}}_6, \ket{T_{3(4)}}_6,\ket{T^*_{3(4)}}_6, $\newline$\ket{T'_{3(4)}}_6, \ket{T'^*_{3(4)}}_6, \ket{T''_{3(4)}}_6$& -2 & 2 & 2 \\ \cline{2-5}
&$\ket{W_{5(6)}}_6, \ket{T_{5(6)}}_6,\ket{T^*_{5(6)}}_6,$\newline$ \ket{T'_{5(6)}}_6, \ket{T'^*_{5(6)}}_6, \ket{T''_{5(6)}}_6$ & 2 & -2 & 2 \\ \cline{2-5}
&$\ket{W_{7}}_6, \ket{T_{7}}_6,\ket{T^*_{7}}_6, $\newline$\ket{T'_{7}}_6, \ket{T'^*_{7}}_6, \ket{T''_{7}}_6$& -2 & 2 & 6 \\ \cline{2-5}
&$\ket{W_{8(9)}}_6, \ket{T_{8(9)}}_6,\ket{T^*_{8(9)}}_6, $\newline$\ket{T'_{8(9)}}_6, \ket{T'^*_{8(9)}}_6, \ket{T''_{8(9)}}_6$& 2 & 2 &-2 \\ \cline{2-5}
\raisebox{19ex}[0pt]{6-qubit}& $\ket{\text{GHZ}'_1}_6, \ket{\text{GHZ}'_2}_6 $& 6 &- &-\\ \cline{1-5}
\end{tabular}
\caption{\label{t1} The energies of basis states of TI for different interactions. }
\end{table}
\end{widetext}

\begin{table}
\center
\begin{tabular}{c|c|c}
\hline
 State & $H_0$ & $H_{nl}=\otimes_{i=1}^N \sigma_i^x$ \\ \hline
 $\ket{0}^{\otimes N}, \ket{1}^{\otimes N}$ & N & 0 \\ \hline
 $\ket{\text{GHZ}_1}_N=\tfrac{1}{\sqrt{2}}\left(\ket{0}^{\otimes N}+\ket{1}^{\otimes N}\right)$& & 1 \\ \cline{1-1} \cline{3-3}
 $\ket{\text{GHZ}_2}_N=\tfrac{1}{\sqrt{2}}\left(\ket{0}^{\otimes N}-\ket{1}^{\otimes N}\right)$& \raisebox{2ex}[0pt]{N} & -1 \\  \hline
\end{tabular}
\caption{\label{t2} Distinguishing fully separable states $\ket{0}^{\otimes N}, \ket{1}^{\otimes N}$ and GHZ-like states.}
\end{table}

}

\section*{Appendix: examples of the  symmetric basis states}

This appendix provides several examples of the complete orthogonal basis of states of TI. Set $\mathcal{T_{R}}_N=\otimes_{n=1}^N\sigma_n^x$ for convenience.

\subsection{3-qubit case}
\begin{eqnarray}
\text{\it 1-period:}&&
\ket{1}_3=\ket{111}; \ket{0}_3=\ket{000}\nonumber\\
\text{\it 3-period:}\nonumber\\
\ket{W_1}_3&=&\frac{1}{\sqrt{3}}\left(\ket{100}+\ket{010}+\ket{001}\right)\nonumber\\
\ket{T_1}_3&=&\frac{1}{\sqrt{3}}\left(\ket{100}+e^{i 2\pi/3}\ket{010}+e^{i 4\pi/3}\ket{001}\right)\nonumber\\
\ket{T^*_1}_3&=&\left(\ket{T_1}_3\right)^*\nonumber\\
\ket{W_2}_3&=&\mathcal{T_R}_3\ket{W_1}_3; \ket{T_2}_3=\mathcal{T_R}_3\ket{T_1}_3; \ket{T^*_2}_3=\left(\ket{T_2}_3\right)^*,
\end{eqnarray}
It is easy to show
\begin{eqnarray}
\hat{T}\ket{T_1}_3&=&e^{- i 2\pi/3}\ket{T_1}_3\nonumber\\
\hat{T}\ket{T_2}_3&=&e^{i 2\pi/3}\ket{T_2}_3
\end{eqnarray}
The orthogonality can be easily proved by $1+e^{i 2\pi/3}+e^{i 4\pi/3}=0$.

It is obvious
\begin{eqnarray}
\ket{T_1}_3=I\otimes \left(\begin{array}{cc}e^{i 2\pi/3} & 0 \\ 0 & 1 \end{array}\right)\otimes\left(\begin{array}{cc}e^{i 4\pi/3} & 0 \\ 0 & 1 \end{array}\right)\ket{W_1}_3.
\end{eqnarray}
A similar relation can be found for $\ket{T_2}_3$.  Thus there are two inequivalent classes $\left\{\ket{1}_3, \ket{0}_3\right\}$ and $\left\{\ket{W_{1(2)}}_3, \ket{T_{1(2)}}_3, \ket{T^*_{1(2)}}_3\right\}$.

\subsection{4-qubit case}
\begin{eqnarray}
\text{\it 1-period:}&&
\ket{1}_4=\ket{1111}; \ket{0}_4=\ket{0000}\nonumber\\
\text{\it 2-period:}\nonumber\\
\ket{\text{GHZ}'_1}_4&=&\frac{1}{\sqrt{2}}\left(\ket{1010}+\ket{0101}\right)\nonumber\\
\ket{\text{GHZ}'_2}_4&=&\frac{1}{\sqrt{2}}\left(\ket{1010}-\ket{0101}\right)\nonumber\\
\text{\it 4-period:}\nonumber\\
\ket{W_1}_4&=&\frac{1}{2}\left(\ket{1000}+\ket{0100}+\ket{0010}+\ket{0001}\right)\nonumber\\
\ket{T_1}_4&=&\frac{1}{2}\left(\ket{1000}+e^{i \pi/2}\ket{0100}\right.\nonumber\\
 &+&\left.e^{i \pi}\ket{0010}+e^{i 3\pi/2}\ket{0001}\right)\nonumber\\
\ket{T^*_1}_4&=&\left(\ket{T_1}_4\right)^*\nonumber\\
\ket{T'_1}_4&=&\frac{1}{2}\left(\ket{1000}-\ket{0100}+\ket{0010}-\ket{0001}\right)\nonumber\\
\ket{W_2}_4&=&\mathcal{T_R}_4\ket{W_1}_4;\ket{T_2}_4=\mathcal{T_R}_4\ket{T_1}_4;\nonumber\\ \ket{T^*_2}_4&=&\left(\ket{T_2}_4\right)^*; \ket{T'_2}_4=\mathcal{T_R}_4\ket{T'_1}_4\nonumber\\
\ket{W_3}_4&=&\frac{1}{2}\left(\ket{1100}+\ket{0110}+\ket{0011}+\ket{1001}\right)\nonumber\\
\ket{T_3}_4&=&\frac{1}{2}\left(\ket{1100}+e^{i \pi/2}\ket{0110}\right.\nonumber\\
&+&\left.e^{i \pi}\ket{0011}+e^{i 3\pi/2}\ket{1001}\right)\nonumber\\
\ket{T^*_3}_4&=&\left(\ket{T_3}_4\right)^*\nonumber\\
\ket{T'_3}_4&=&\frac{1}{2}\left(\ket{1100}-\ket{0110}+\ket{0011}-\ket{1001}\right)
\end{eqnarray}
It is easy to verify
\begin{eqnarray}
\hat{T}\ket{\text{GHZ}'_{1(2)}}_4&=&\pm\ket{\text{GHZ}'_{1(2)}}_4\nonumber\\
\hat{T}\ket{T_{1(2,3)}}_4&=&-i\ket{T_{1(2,3)}}_4
\end{eqnarray}
Obviously one has SLOCC equivalences, for example,
\begin{eqnarray}
\ket{T_1}_4&=&I\otimes \left(\begin{array}{cc}e^{i \pi/2} & 0 \\ 0 & 1 \end{array}\right)\otimes\left(\begin{array}{cc}e^{i \pi} & 0 \\ 0 & 1 \end{array}\right)\otimes\nonumber\\&&\left(\begin{array}{cc}e^{i 3\pi/2} & 0 \\ 0 & 1 \end{array}\right)\ket{W_1}_4\nonumber\\
\ket{T'_1}_4&=&I\otimes \left(\begin{array}{cc}-1 & 0 \\ 0 & 1 \end{array}\right)\otimes I\otimes\left(\begin{array}{cc}-1 & 0 \\ 0 & 1 \end{array}\right)\ket{W_1}_4
\end{eqnarray}
With respect to distinct cyclic units, there are four inequivalent classes,
$\left\{\ket{1}_4, \ket{0}_4\right\}$, $\left\{\ket{\text{GHZ}'_1}_4, \ket{\text{GHZ}'_2}_4\right\}$,$\left\{\ket{W_{1(2)}}_4, \ket{T_{1(2)}}_4,\ket{T^*_{1(2)}}_4, \ket{T'_{1(2)}}_4\right\}$  and $\left\{\ket{W_3}_4, \ket{T_3}_4, \ket{T^*_3}_4, \ket{T'_3}_4\right\}$.

\subsection{5-qubit case}

\begin{eqnarray}
\text{\it 1-period:}&&
\ket{1}_5=\ket{11111}; \ket{0}_5=\ket{00000}\nonumber\\
\text{\it 5-period:}&&\nonumber\\
\ket{W_1}_5&=&\frac{1}{\sqrt{5}}\left(\ket{10000}+\ket{01000}+\ket{00100}\right.\nonumber\\ &&\left.+\ket{00010}+\ket{00001}\right)\nonumber\\
\ket{T_1}_5&=&\frac{1}{\sqrt{5}}\left(\ket{10000}+e^{i 2\pi/5}\ket{01000}+e^{i 4\pi/5}\ket{00100}\right.\nonumber\\ &&\left.+e^{i 6\pi/5}\ket{00010}+e^{i 8\pi/5}\ket{00001}\right)\nonumber\\
\ket{T^*_1}_5&=&\left(\ket{T_1}_5\right)^*\nonumber\\
\ket{T'_1}_5&=&\frac{1}{\sqrt{5}}\left(\ket{10000}+e^{i 4\pi/5}\ket{01000}+e^{i 8\pi/5}\ket{00100}\right.\nonumber\\ &&\left.+e^{i 12\pi/5}\ket{00010}+e^{i 16\pi/5}\ket{00001}\right)\nonumber\\
\ket{T'^*_1}_5&=&\left(\ket{T'_1}_5\right)^*\nonumber\\
\ket{W_2}_5&=&\frac{1}{\sqrt{5}}\left(\ket{01111}+\ket{10111}+\ket{11011}\right.\nonumber\\ &&\left.+\ket{11101}+\ket{11110}\right)\nonumber\\
\ket{T_2}_5&=&\frac{1}{\sqrt{5}}\left(\ket{01111}+e^{i 2\pi/5}\ket{10111}+e^{i 4\pi/5}\ket{11011}\right.\nonumber\\ &&\left.+e^{i 6\pi/5}\ket{11101}+e^{i 8\pi/5}\ket{11110}\right)\nonumber\\
\ket{T^*_2}_5&=&\left(\ket{T_2}_5\right)^*\nonumber\\
\ket{T'_2}_5&=&\frac{1}{\sqrt{5}}\left(\ket{01111}+e^{i 4\pi/5}\ket{10111}+e^{i 8\pi/5}\ket{11011}\right.\nonumber\\ &&\left.+e^{i 12\pi/5}\ket{11101}+e^{i 16\pi/5}\ket{11110}\right)\nonumber\\
\ket{T'^*_2}_5&=&\left(\ket{T'_2}_5\right)^*.
\end{eqnarray}
For the other basis states  not listed here, we present the corresponding cyclic units only: $\{11000\}$ and $\{00111\}$ for the series of $\{\ket{W_3}_5\}$ and $\{\ket{W_4}_5\}$, $\{10100\}$ and $\{01011\}$ for the series of $\{\ket{W_5}_5\}$ and $\{\ket{W_6}_5\}$.

Then there are 4 classes, dependent on the periodicity and the cyclic unit,
\begin{eqnarray}
&&\left\{\ket{\text{GHZ}_{1(2)}}_5 \right\},\nonumber\\
&&\left\{\ket{W_{1(2)}}_5, \ket{T_{1(2)}}_5,\ket{T^*_{1(2)}}_5 \ket{T'_{1(2)}}_5,  \ket{T^{'*}_{1(2)}}_5 \right\},\nonumber\\
&&\left\{\ket{W_{3(4)}}_5, \ket{T_{3(4)}}_5,\ket{T^*_{3(4)}}_5, \ket{T'_{3(4)}}_5,\ket{T^{'*}_{3(4)}}_5 \right\},\nonumber\\
&&\left\{\ket{W_{5(6)}}_5, \ket{T_{5(6)}}_5,\ket{T^*_{5(6)}}_5, \ket{T'_{5(6)}}_5,\ket{T^{'*}_{5(6)}}_5 \right\}
\end{eqnarray}

\subsection{6-qubit case}

\begin{eqnarray}
\text{\it 1-period:}&&
\ket{1}_6=\ket{111111}; \ket{0}_6=\ket{000000}\nonumber\\
\text{\it 2-period:}&&\nonumber\\
\ket{\text{GHZ}'_1}_6&=&\frac{1}{\sqrt{2}}\left(\ket{101010}+\ket{010101}\right)\nonumber\\
\ket{\text{GHZ}'_2}_6&=&\frac{1}{\sqrt{2}}\left(\ket{101010}-\ket{010101}\right)\nonumber
\end{eqnarray}
\begin{eqnarray}
\text{\it 3-period:}&&\nonumber\\
\ket{W_{0}}_6&=&\frac{1}{\sqrt{3}}\left(\ket{100100}+\ket{010010}+\ket{001001}\right)\nonumber\\
\ket{T_{0}}_6&=&\frac{1}{\sqrt{3}}\left(\ket{100100}+e^{i 2\pi/3}\ket{010010}\right.\nonumber\\
&+&\left.e^{i 4\pi/3}\ket{001001}\right)\nonumber\\
\ket{T^*_{0}}_6&=&\left(\ket{T_{0}}_6\right)^*\nonumber\\
\ket{W'_{0}}_6&=&\mathcal{T_R}_6\ket{W_{0}}_6; \ket{T'_{0}}_6=\mathcal{T_R}_6\ket{T_{0}}_6; \ket{T'^*_{0}}_6=\left(\ket{T'_{0}}_6\right)^*\nonumber\\
\text{\it 6-period:}&&\nonumber\\
\ket{W_1}_6&=&\frac{1}{\sqrt{6}}\left(\ket{100000}+\ket{010000}+\ket{001000}\right.\nonumber\\ &+&\left.\ket{000100}+\ket{000010}+\ket{000001}\right)\nonumber\\
\ket{T_1}_6&=&\frac{1}{\sqrt{6}}\left(\ket{100000}+e^{i \pi/3}\ket{010000}\right.\nonumber\\
&+&\left.e^{i 2\pi/3}\ket{001000}+e^{i \pi}\ket{000100}\right.\nonumber\\
&+&\left.e^{i 4\pi/3}\ket{000010}+e^{i 5\pi/3}\ket{000001}\right).\nonumber\\
\ket{T'_1}_6&=&\frac{1}{\sqrt{6}}\left(\ket{100000}+e^{i 2\pi/3}\ket{010000}\right.\nonumber\\
&+&\left.e^{i 4\pi/3}\ket{001000}+\ket{000100}\right.\nonumber\\
&+&\left.e^{i 2\pi/3}\ket{000010}+e^{i 4\pi/3}\ket{000001}\right)\nonumber\\
\ket{T^*_1}_6&=&\left(\ket{T_1}_6\right)^*; \ket{T'^*_1}_6=\left(\ket{T'_1}_6\right)^*\nonumber\\
\ket{T''_1}_6&=&\frac{1}{\sqrt{6}}\left(\ket{100000}-\ket{010000}+\ket{001000}\right.\nonumber\\ &-&\left.\ket{000100}+\ket{000010}-\ket{000001}\right)\nonumber\\
\ket{W_2}_6&=&\mathcal{T_R}_6\ket{W_1}_6; \ket{T_2}_6=\mathcal{T_R}_6\ket{T_1}_6\nonumber\\
\ket{T^*_2}_6&=&\left(\ket{T_2}_6\right)^*; \ket{T'_2}_6=\mathcal{T_R}_6\ket{T'_1}_6\nonumber\\
\ket{T'^*_2}_6&=&\left(\ket{T'_2}_6\right)^*; \ket{T''_2}_6=\mathcal{T_R}_6\ket{T''_1}_6.
\end{eqnarray}
The other basis states are not listed here. Instead we present the corresponding cyclic units, by which they can be obtained in the same way as above. For C5 (see the next paragraph), cyclic units are "$110000$" and "$001111$". Similarly, "$101000$" and "$010111$" for C6; "$111000$" for C7; "$101100$" and "$110100$" for C8.

All basis states can be divided into 8 classes,
\begin{eqnarray}
&&C1: \left\{\ket{1}_6, \ket{0}_6\right\},\nonumber\\
&&C2:\left\{\ket{\text{GHZ}'_1}_6, \ket{\text{GHZ}'_2}_6\right\}\nonumber\\
&&C3:\left\{\ket{W^{(\prime)}_{0}}_6, \ket{T^{(\prime)}_{0}}_6,\ket{T^{(\prime)*}_{0}}_6\right\}\nonumber\\
&&C4:\left\{\ket{W_{1(2)}}_6, \ket{T_{1(2)}}_6,\ket{T^*_{1(2)}}_6, \ket{T'_{1(2)}}_6, \ket{T'^*_{1(2)}}_6, \ket{T''_{1(2)}}_6\right\}\nonumber\\
&&C5:\left\{\ket{W_{3(4)}}_6, \ket{T_{3(4)}}_6,\ket{T^*_{3(4)}}_6, \ket{T'_{3(4)}}_6, \ket{T'^*_{3(4)}}_6, \ket{T''_{3(4)}}_6 \right\}\nonumber\\
&&C6:\left\{ \ket{W_{5(6)}}_6, \ket{T_{5(6)}}_6,\ket{T^*_{5(6)}}_6, \ket{T'_{5(6)}}_6, \ket{T'^*_{5(6)}}_6, \ket{T''_{5(6)}}_6\right\}\nonumber\\
&&C7:\left\{ \ket{W_{7}}_6, \ket{T_{7}}_6,\ket{T^*_{7}}_6, \ket{T'_{7}}_6, \ket{T'^*_{7}}_6, \ket{T''_{7}}_6\right\}\nonumber\\
&&C8:\left\{ \ket{W_{8(9)}}_6, \ket{T_{8(9)}}_6,\ket{T^*_{8(9)}}_6, \ket{T'_{8(9)}}_6, \ket{T'^*_{8(9)}}_6, \ket{T''_{8(9)}}_6\right\}
\end{eqnarray}

\end{document}